\documentclass{jcmlatex}
\usepackage{amssymb}
\usepackage{graphicx}

\def\JCMvol{xx}
\def\JCMno{x}
\def\JCMyear{200x}
\def\JCMreceived{xxx}
\def\JCMrevised{xxx}
\def\JCMaccepted{xxx}

\begin{document}

\markboth{L. FERN\'ANDEZ-JAMBRINA AND F. P\'EREZ-ARRIBAS}
         {Developable surface patches bounded by NURBS curves}

\title{DEVELOPABLE SURFACE PATCHES BOUNDED BY NURBS CURVES}

\author{Leonardo Fern\'andez-Jambrina
\thanks{ETSI Navales, Universidad Polit\'ecnica de Madrid, 
28040-Madrid, Spain\\
  Email: leonardo.fernandez@upm.es
          } and Francisco P\'erez-Arribas\thanks{ETSI Navales, 
		  Universidad Polit\'ecnica de Madrid, 
28040-Madrid, Spain\\
  Email: francisco.perez.arribas@upm.es
          }
    }

\maketitle

\begin{abstract}
In this paper we construct developable surface patches which are
bounded by two rational or NURBS curves, though the resulting patch is
not a rational or NURBS surface in general.  This is accomplished by
reparameterizing one of the boundary curves.  The reparameterization
function is the solution of an algebraic equation.  For the relevant
case of cubic or cubic spline curves, this equation is quartic at
most, quadratic if the curves are B\'ezier or splines and lie on
parallel planes, and hence it may be solved either by standard
analytical or numerical methods.
\end{abstract}



\section{Introduction\label{introd}}

Developable surfaces play an important role in differential geometry 
as surfaces with vanishing Gaussian curvature. From the point of view of 
intrinsic geometry, developable surfaces cannot be distinguished from 
the plane. Only when they are embedded in three-dimensional space, 
different surfaces arise. The embedding of the planar surface in space 
has to preserve lengths and angles between curves. Metric 
properties are not altered and hence the planar surface may be cut or 
folded, but not stretched or deformed.

On the other hand, these geometrical properties are of relevance for 
industry. In textile design one starts with a planar piece of cloth to 
produce garments and their quality improves if the cloth is not 
stretched. In naval industry one has to adapt planar sheets of steel  
to the form of the hull of a vessel. This can be done with a folding 
machine if the result is a developable surface, avoiding the 
application of heat and reducing the costs. They are also useful for 
modeling pages of a book \cite{bartoli} for 3D reconstruction and 
they can also be found in architectural constructions \cite{architecture}. 

The main problem for addressing developable surfaces in Geometrical 
Design is that the null Gaussian curvature requirement takes the form 
of a non-linear equation when expressed in terms of the vertices of 
the control net of the surface. 

This issue has been handled in several ways. A thorough review may be 
found in \cite{computational-line}. In \cite{lang} rational
B\'ezier surfaces are addressed and the null Gaussian curvature
condition is solved explicitly for low degrees.  $C^2$-spline
developable surfaces are constructed in \cite{maekawa}.  Another
restriction is considering boundary curves for the developable surface
on parallel planes as in \cite{aumann0} and \cite{frey}. A different point of view relies on solving the null Gaussian
curvature in the dual space of planes \cite{ravani, pottmann-farin, wallner}.

Concerning the applications in industry, quasi-developable surfaces
are constructed in \cite{chalfant} and \cite{arribas}.  In
\cite{kilgore} developable surfaces for designing ship hulls are
constructed by graphical methods.  Developable surfaces can also be
approximated with spline cones as in \cite{leopoldseder}.  A different
and successful approach for approximate developable surfaces bounded
by polylines, grounded on convex hulls, is shown in \cite{rose}, with
examples for garments.

Application of the de Casteljau algorithm has lead to several fruitful
approaches as in \cite{sequin}.  In \cite{aumann} a family of
developable surfaces is constructed through a B\'ezier curve of
arbitrary degree. This is useful for solving interpolation problems
\cite{aumann1}.  These results have been extended to spline curves of
arbitrary degree in \cite{leonardo-developable, leonardo-elevation}
and to B\'ezier triangular surfaces \cite{leonardo-triangle}. In 
\cite{geomdevelop} developable surfaces with several patches linked 
with $G^{1}$-continuity are constructed.

In \cite{origami} the non-linear conditions are expressed as quadratic
equations and this is used to devise a constraint for interactive modeling.

Finally, in \cite{leonardo-bezier} it is shown that the developable 
surfaces which can be constructed with Aumann's algorithm are the 
ones with a polynomial edge of regression. This poses an interesting problem. When we interpolate a ruled 
surface between two parameterized curves, $c(t)$ and $d(t)$, besides 
the obvious way,
\[b(t,v)=(1-v)c(t)+vd(t),\qquad v\in[0,1],\]
there are other infinite possibilities, depending on the choice of 
parameterizations for the bounding curves. In this paper we focus on 
this issue.

The paper is organised as follows: In Section~\ref{develop} we
introduce developable surface patches bounded by rational B\'ezier
curves of arbitrary degree $n$.  We look for the most general solution
to this problem by reparameterizing one of the curves.  The
reparameterization function is shown to satisfy an algebraic equation
of degree $2n-2$ at most, or of degree $n-1$ if the bounding curves
are polynomial and lie on parallel planes.  Examples are provided in Section~\ref{examp}.
In Section~\ref{devspline} it is shown how the results can be applied
to developable surface patches bounded by NURBS curves.  A final
section of conclusions is included.

\section{Developable patches bounded by rational curves\label{develop}}

We start with two rational curves of degree $n$, $c(t)$, $d(T)$,
$t,T\in[0,1]$ and respective control polygons
$\{c_{0},\ldots,c_{n}\}$, $\{d_{0},\ldots,d_{n}\}$ and lists of 
weights $\{w_{0},\ldots,w_{n}\}$, $\{\omega_{0},\ldots,\omega_{n}\}$.  We may think of
$T=T(t)$ as a function of $t$ in order to construct a parameterized
ruled surface,
\[b(t,v)=(1-v)c(t)+vd(T(t))=(1-v)c(t)+v\hat d(t),\qquad t,v\in[0,1],\]
denoting the reparameterized curve as $\hat d(t):=d(T(t))$. We shall denote by a comma the derivative with respect to $t$ and by a dot 
the derivative with respect to $T$. 

A normal vector $N(t,v)$ to the surface at $b(t,v)$ may be calculated,
\begin{eqnarray}N(t,v)&:=&b_{t}(t,v)\times b_{v}(t,v)=\left((1-v)c'(t)+v\hat 
d'(t)\right)\times \left(\hat d(t)-c(t)\right)\nonumber\\&=&
(1-v)N(t,0)+vN(t,1),\label{normal}\end{eqnarray}
as a barycentric combination of the normal $N(t,0)$ at $c'(t)$ and 
the normal  $N(t,1)$ at $\hat d'(t)$.

\begin{figure}
\begin{center}
    \includegraphics[height=0.15\textheight]{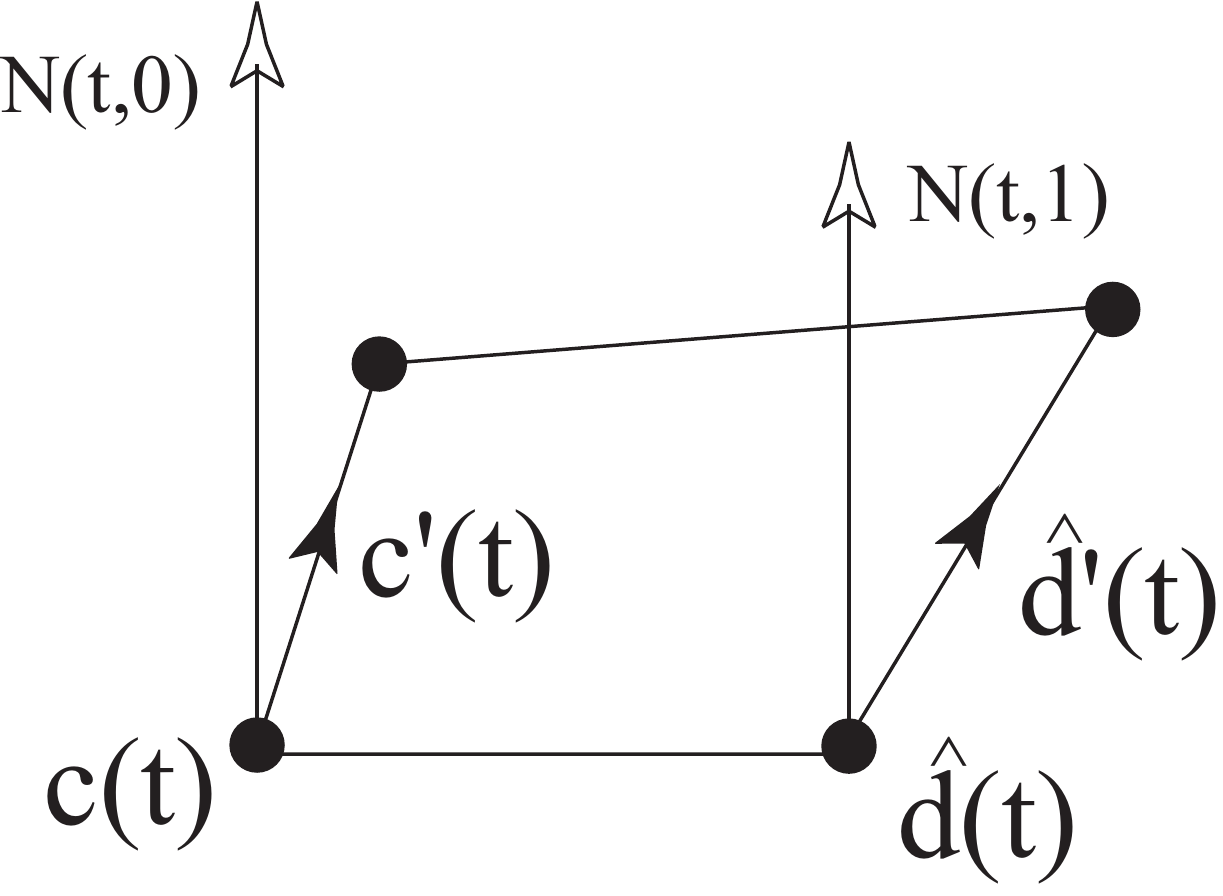}
\end{center}
\caption{Normal vectors to a developable surface at the ends of a 
ruling\label{normalv}}
\end{figure}

In the case of developable surfaces \cite{struik}, $N(t,0)$ and $N(t,1)$ are
parallel for all values of $t$ (See 
Fig.~\ref{normalv}). In order to avoid singular points for which $N(t,v)$ is a zero vector, we 
require that $N(t,0)$ and $N(t,1)$ point to the same side of the 
tangent plane along the ruling corresponding to $t$. Otherwise, the 
vector $N(t,v)$  would vanish for a value $v\in (0,1)$.

The condition for the parameterized surface to be developable 
\cite{struik} is coplanarity of 
$c'(t)$, $\hat d'(t)$, $\hat d(t)-c(t)$ for all values of $t$,
\[ \det\left(c'(t), \hat d'(t), \hat d(t)-c(t)\right)= 0.\] 

By the chain rule,
\[\hat d'(t)=T'(t)\dot d(T)|_{T=T(t)},\]
and multilinearity of the determinant, the previous condition
is equivalent to 
\begin{equation}\label{algdev} \det\left(c'(t), \dot d(T), 
d(T)-c(t)\right)= 0,\end{equation}
which is expected to be satisfied by a function $T(t)$.

We may think of the previous expression as the equation for the
reparameterizations $T(t)$ that allows us to construct a developable
surface bounded by the curves $c$, $d$.  The main advantage of this
form of the equation is that it is purely algebraic instead of
differential and that it
depends on a single function $T(t)$.

We may impose from the beginning that $c_{0},c_{1},d_{0},d_{1}$ be 
coplanar in order to fulfill the developability condition at $t=0$. 
Similarly, $c_{n},c_{n-1},d_{n},d_{n-1}$ are to be coplanar in order to 
satisfy the developability condition at $t=1$.

\begin{figure}
\begin{center}
    \includegraphics[height=0.15\textheight]{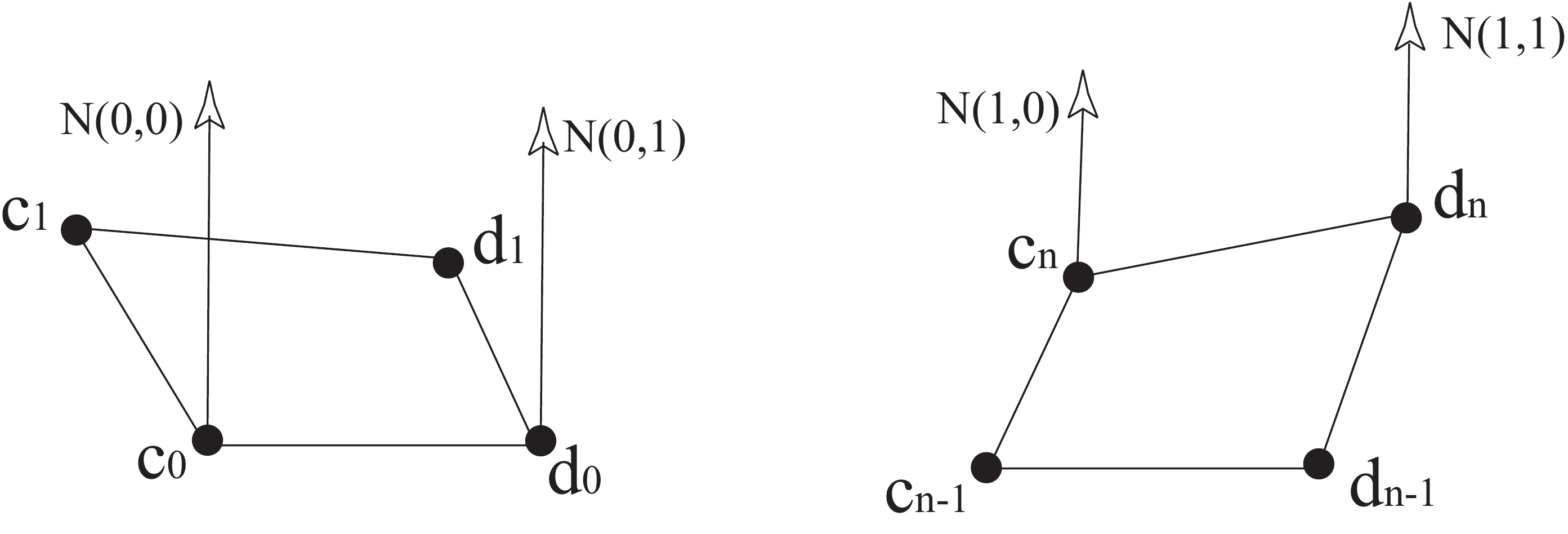}
\end{center}
\caption{Normal vectors to a developable surface at the first 
ruling\label{normalv1}}
\end{figure}

We also require that $N(0,0)=n(c_{1}-c_{0})\times (d_{0}-c_{0})$ 
and $N(0,1)=n(d_{1}-d_{0})\times (d_{0}-c_{0})$ point to the 
same side of the tangent plane at $t=0$  and that 
$N(1,0)=n(c_{n}-c_{n-1})\times (d_{n}-c_{n})$ 
and $N(1,1)=n(d_{n}-d_{n-1})\times (d_{n}-c_{n})$ point to the 
same side of the tangent plane at $t=1$ (See 
Fig.~\ref{normalv1}). 

If we write $c(t)=\vec{p}(t)/w(t)$, $d(t)=\vec{q}(t)/\omega(t)$, by
explicitly splitting the parameterizations of the curves in their
respective denominators $w(t)$, $\omega(t)$ and vector numerators
$\vec{p}(t)$, $\vec{q}(t)$ of degree $n$, we may expand (\ref{algdev}) 
as
\begin{eqnarray*}0&=&\det\left(c'(t), \dot d(T), d(T)-c(t)\right)\\&=&
\frac{\det\left(\vec{p}(t),\vec{p'}(t),\omega(T)\dot{\vec{q}}(T)-
\dot\omega(T)q(T)\right)+
\det\left(\vec{q}(T),\dot{\vec{q}}(T),w(t){\vec{p'}}(t)-
w'(t)p(t)\right)}{w(t)^{2}\omega(T)^{2}}.\end{eqnarray*}

In principle the numerator of this expressions provides an equation of degree $2n-1$ in 
either $T$ or $t$, but it is easily seen that the $2n-1$ derivative
with respect to $T$ is identically null and hence
 (\ref{algdev}) is an equation of degree $2n-2$ at most.  For instance, for
the relevant case of cubic curves, the equation is quartic.

Furthermore, if both $c$, $d$ are B\'ezier plane curves and lie in parallel planes, the reduction is 
more dramatic. Denoting by $n)$ the $n$-th derivative, in the $n-1$ 
derivative of the left-hand side of (\ref{algdev}),
\[\det\left(c'(t), d^{n)}(T), d(T)-c(t)\right)= 0\] all terms are coplanar except for $d(T)-c(t)$.
Hence, the $n$-th derivative vanishes identically and (\ref{algdev})
is an equation of degree $n-1$ in the case of curves in parallel
planes. This case is relevant since in many cases one is interested 
in interpolating a surface between plane sections, suchs as stations or 
waterlines of a hull.

\begin{proposition} Let $c(t)$, $d(T)$, $t,T\in[0,1]$ be rational
curves of degree $n$ with coplanar sets of vertices
$c_{0},d_{0},c_{1}, d_{1}$ and $c_{n-1},d_{n-1},c_{n},d_{n}$.

We also require that
$(c_{1}-c_{0})\times (d_{0}-c_{0})$ 
and $(d_{1}-d_{0})\times (d_{0}-c_{0})$ point to the 
same side of the tangent plane at $t=0$  and that 
$(c_{n}-c_{n-1})\times (d_{n}-c_{n})$ 
and $(d_{n}-d_{n-1})\times (d_{n}-c_{n})$  point to the 
same side of the tangent plane at $t=1$. 

The parameterized ruled surface,
\[b(t,v)=(1-v)c(t)+vd(T(t)),\qquad t,v\in[0,1],\]
is a developable surface if the reparameterization function $T(t)$ 
satisfies the algebraic equation
\[\det\left(c'(t), \dot d(T), 
d(T)-c(t)\right)_{T=T(t)}= 0,\]
and is a real monotonically increasing function of $t$.

This equation is of degree $2n-2$ at most.  If both curves are
polynomial and lie in parallel planes, the equation is of degree $n-1$
at most.
\end{proposition}

Among the solutions of (\ref{algdev}) we have to choose the ones 
which are real and monotonically increasing. 
If $T(t)$ is not monotonically increasing, the singular edge of
regression of the developable surface is crossed by the surface patch.  This can be dealt
with using the multiconic development \cite{arribas}, also known as 
Rabl's method \cite{rabl}, \cite{arribas}, modifying the
boundary curves and replacing the region of the surface where $T(t)$
decreases.

We obtain information about this issue, taking the derivative of (\ref{algdev}) 
with respect to $t$,
\begin{eqnarray*}
0&=&\det\left(c''(t), \dot d(T), 
d(T)-c(t)\right)+ \det\left(c'(t), 
\ddot d(T)T'(t), 
d(T)-c(t)\right)\\&+&\left. \det\left(c'(t), \dot d(T), 
\dot d(T)T'(t)-c'(t)\right)\right|_{T=T(t)}.\end{eqnarray*}

Since the last term is trivially zero, we can get an expression for 
the derivative of the reparameterization function,
\begin{equation}\label{monotone}
T'(t)=\left.\frac{\det\left(c''(t), \dot d(T), 
d(T)-c(t)\right)}{\det\left( \ddot d(T),c'(t), 
d(T)-c(t)\right)}\right|_{T=T(t)},\end{equation}
which has to be positive in order to have a monotonically increasing 
reparameterization function $T(t)$.

If $c(t)$ and $d(T)$ are parameterizations of class $C^k$ of 
differentiability, we see that $T(t)$ is at least of class $C^{k-1}$. This 
is relevant for dealing with spline parameterizations. If $c(t)$, 
$d(T)$ are spline curves of degree $n$ without repeated knots, that 
is, of class $C^{n-1}$, the reparameterization function is of class 
$C^{n-2}$.

This expression allows several interpretations.  Since for a
reparameterization function $T(t)$ both $\dot d(T)\times \left.
\left(d(T)-c(t)\right)\right|_{T=T(t)}$ and $\left.  c'(t)\times
\left( d(T)-c(t)\right)\right|_{T=T(t)}$ are normal vectors to the
developable surface along the ruling linking $d(T(t))$ and $c(t)$,
both are parallel to the unitary normal $\nu(t)$ to the developable
surface at this ruling.  We assume here that both normal vectors point
to the same side of the tangent plane along the ruling.  Otherwise,
the developable surface patch would intersect the singular edge of
regression.

\begin{figure}
\begin{center}
    \includegraphics[height=0.15\textheight]{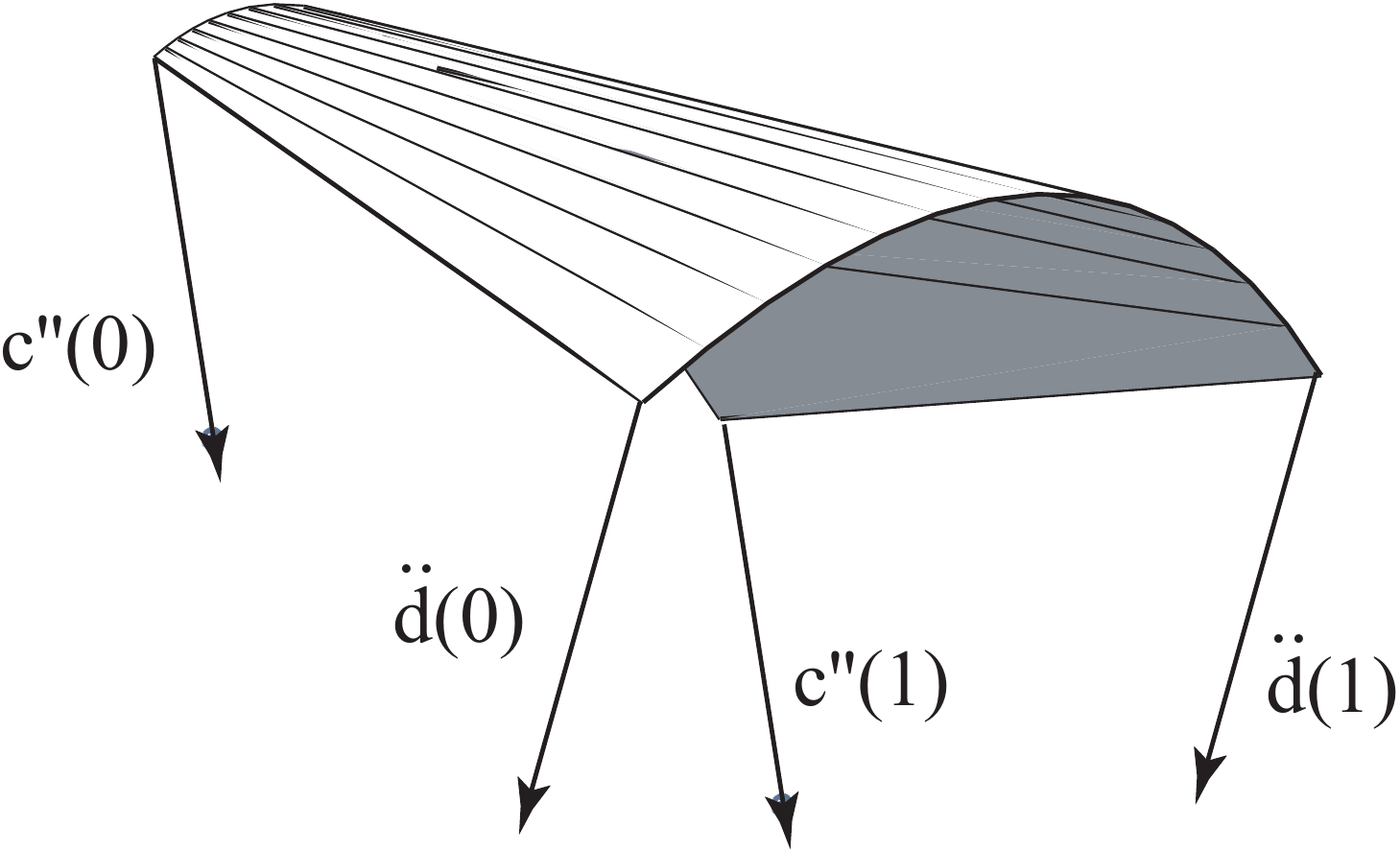}
\end{center}
\caption{Acceleration vectors must point to the same side of the 
tangent plane to the rulings\label{accpar}}
\end{figure}

This means that $T(t)$ is monotonically
increasing if the projections of the accelerations of both curves,
$c''(t)\cdot \nu(t)$, $\left.\ddot d(T)\right|_{T=T(t)}\cdot \nu(t)$, 
have the same sign. That is, if $c''(t)$, $\left.\ddot d(T)\right|_{T=T(t)}$ point to the 
same side of the tangent plane at $t$. In the case of parabolas, for 
instance, these two vectors are respectively parallel to the axes of 
the parabolas (See Fig.~\ref{accpar}).

We see what happens when such requirement is not satisfied:

\begin{example}Developable surface patch bounded by 
parabolas $c$ and $d$ with respective control polygons 
\[\{(0,0,0),(0,1,0),(2,1,0)\},\qquad \{(0,0,1),(0,-3/2,1),(1,-3/2,1)
\}.\]\end{example}

\begin{figure}
\begin{center}
    \includegraphics[height=0.2\textheight]{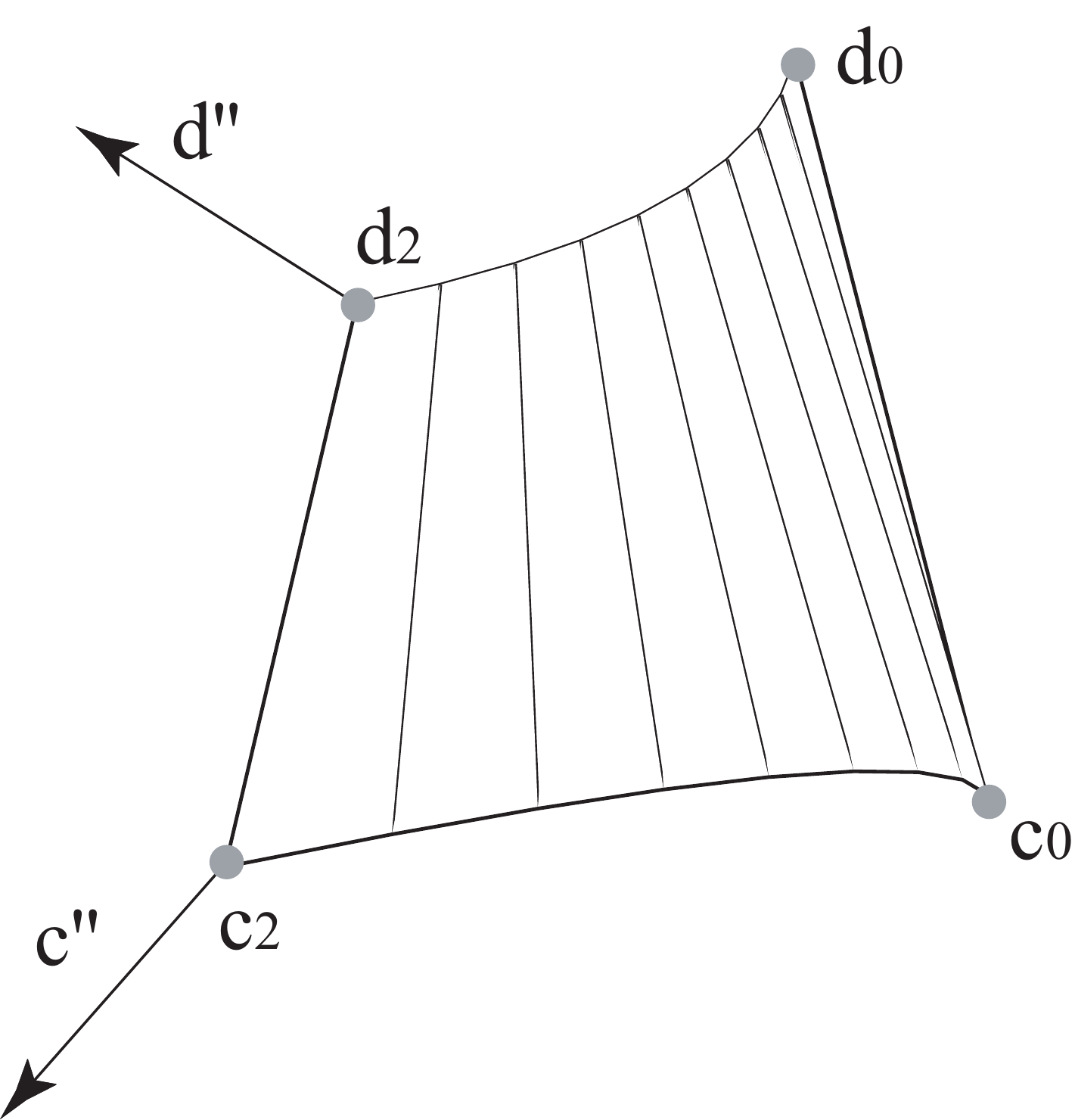}
\end{center}
\caption{Ruled surface patch bounded by parabolas on parallel 
planes\label{badparabola}}
\end{figure}

The parameterizations of these curves,\[ c(t)=(2 t^{2} , -t^{2 } + 2 t, 0), 
\qquad d(T)=(T^{2} , 3T^{2}/2  - 3 T, 1),\]
require a reparameterization \[T(t)=\frac{3t}{4t-1},\]
which is not a growing function. This happens because the 
accelerations of the parabolas, $c''$, $d''$, point to opposite sides, as it can be 
seen in Fig.~\ref{badparabola}.

The acceleration of a parameterized curve may be split into tangential
acceleration and normal acceleration in a standard way.  The
tangential part does not contribute to $c''(t)\cdot \nu(t)$.  

Hence, we may replace $c''(t)$ by the normal acceleration or curvature
vector $k_{c}(t)$ of the curve $c(t)$ in the expression $c''(t)\cdot
\nu(t)$.  And also we may replace $\ddot d(T)$ by $k_{d}(T)$.  But the
projection of the curvature vectors of the curves along the unitary
normal to the surface \cite{struik} are their normal curvatures,
$k_{n,c}$, $k_{n,d}$,
\[k_{n,c}=k_{c}\cdot\nu,\qquad k_{n,d}=k_{d}\cdot\nu,\]
and we can summarise this result in the following way:

\begin{proposition}Let $c(t)$, $d(T)$, $t,T\in[0,1]$ be parameterized
curves with coplanar sets of vertices $c_{0},d_{0},c_{1},d_{1}$ 
and $c_{n-1},d_{n-1},c_{n},d_{n}$. 

We also require that
$(c_{1}-c_{0})\times (d_{0}-c_{0})$ 
and $(d_{1}-d_{0})\times (d_{0}-c_{0})$ point to the 
same side of the tangent plane at $t=0$  and that 
$(c_{n}-c_{n-1})\times (d_{n}-c_{n})$ 
and $(d_{n}-d_{n-1})\times (d_{n}-c_{n})$  point to the 
same side of the tangent plane at $t=1$.

Let $T(t)$ be a reparameterization function so that 
\[b(t,v)=(1-v)c(t)+vd(T(t)),\qquad t,v\in[0,1],\]
is a developable surface. 
$T(t)$ is a monotonically increasing function if  and only if for all 
$t$,
\[\mathrm{sgn}\left(c''(t)\cdot\nu(t)\right)=
\mathrm{sgn}\left.\left(\ddot 
d(T)\cdot\nu(t)\right)\right|_{T=T(t)},\] where $\nu(t)$ is the 
unitary normal to the surface along the ruling at $t$.

Or equivalently, for the normal curvatures $k_{n,c}$, $k_{n,d}$ of 
both curves\[\mathrm{sgn}\left(k_{n,c}(t)\right)=
\mathrm{sgn}\left(k_{n,d}(T(t)\right),\] for all values of $t$. 
In the case of parameterizations of class $C^k$ of differentiability, $T(t)$
is of class $C^{k-1}$.
\end{proposition}


In case we have a developable surface patch with a region where the
reparameterization $T(t)$ is not a growing function in an interval
$[t_{i},t_{f}]$, ($[T_{i},T_{f}]$ for the curve $d(T)$), we have a
regression area where the rulings overlap as in Fig~\ref{regres}.

We may amend this flaw using the multiconic development 
\cite{rabl, arribas}, replacing the regression area by pieces of 
cones and modifying the curve $d(T)$.

First, we obtain two sequences of equally spaced points,
$c_{0},\ldots,c_{r}$; $d_0,\ldots,d_{r}$, on the curves $c(t)$, 
$d(T)$, such that $c_{0}=c(t_{i})$, $c_{r}=c(t_{f})$, 
$d_{0}=d(T_{i})$, $d_{r}=d(T_{f})$, which provide lines 
$L_{0},\ldots, L_{r}$ linking pairs of points on the curves.

The line $L_{1}$ linking $c_{1}$ and $d_{1}$ does not belong to a
developable surface, but we replace it by a ruling of a cone. With 
this aim, we define the plane $\alpha_{1}$ containing the line 
$L_{1}$ and which is orthogonal to the plane $\gamma_{1}$, containing 
the same line $L_{1}$ and the velocity $d'_{1}$ of the curve $d(t)$ 
at $d_{1}$. 

The plane $\alpha_{1}$ meets the ruling $L_{0}$ at the 
point $a_{1}$, which is be the vertex of the new cone. The new 
ruling replacing $L_{1}$ is $\tilde L_{1}$, through $a_{1}$ and 
$c_{1}$. 

The endpoint of $\tilde L_{1}$ is the point $\tilde d_{1}$ 
where this new ruling meets the osculating plane $\beta_{1}$ of the 
curve $d(t)$ at $d_{1}$ (the plane with more contact with the curve 
$d(t)$ at $d_{1}$, since it contains $d_{1}$, $d'_{1}$ and $d''_{1}$).

After obtaining the new ruling $\tilde L_{1}$ and its endpoint 
$\tilde d_{1}$ we proceed to compute the ruling $\tilde L_{2}$ 
replacing $L_{2}$ by constructing a cone with $\tilde L_{1}$ in a 
similar fashion. The procedure finishes after recalculating the new 
rulings through $c_{1},\ldots,c_{r}$.
\begin{figure}
\begin{center}
    \includegraphics[height=0.3\textheight]{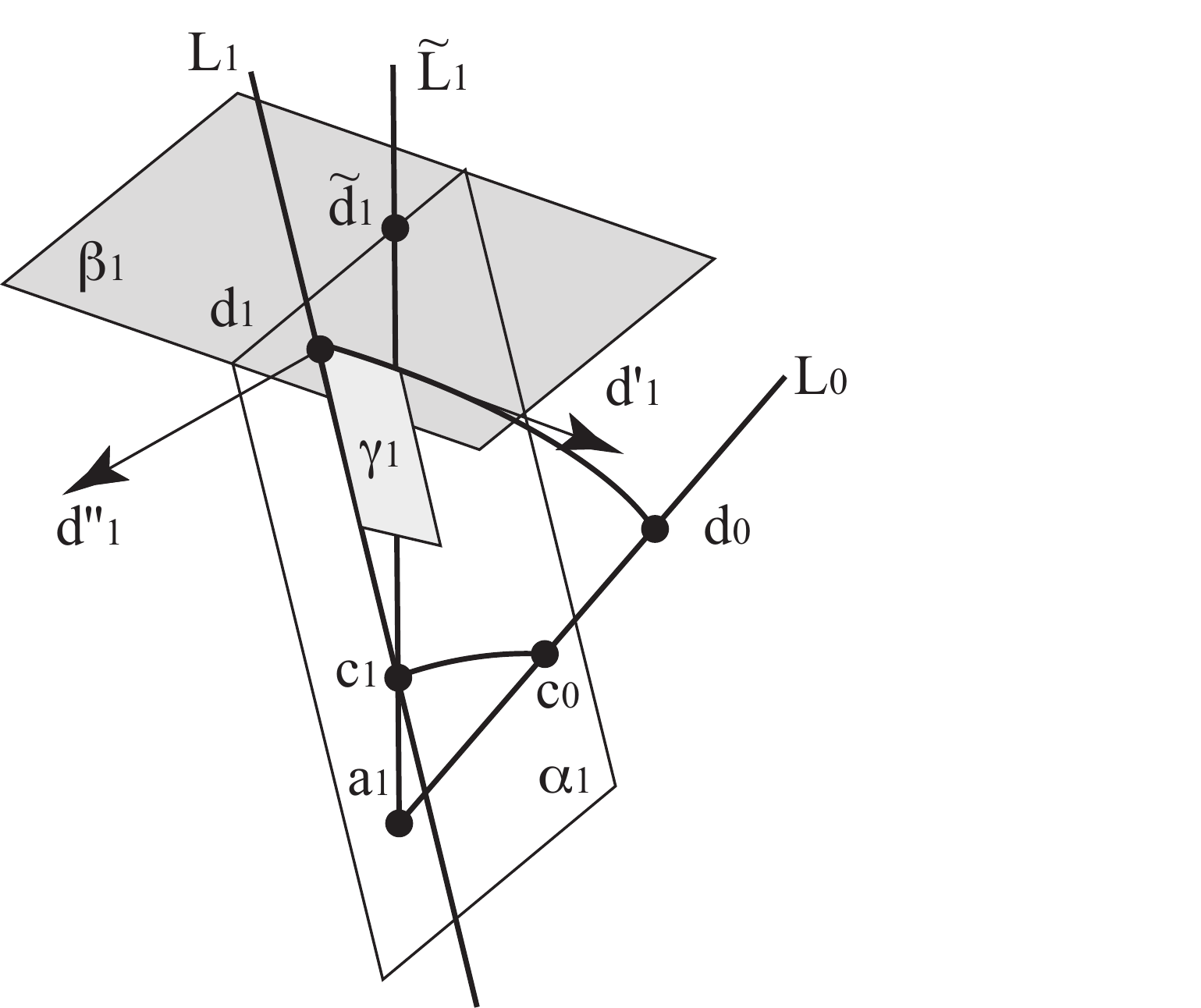}
\end{center}
\caption{Replacement of the ruling $L_{1}$ by $\tilde L_{1}$ in the 
multiconic development\label{multiconic}}
\end{figure}

\section{Examples\label{examp}}

The procedure for constructing developable surfaces bounded by two 
rational B\'ezier curves $c(t)$ and $d(T)$ satisfying the requirements of 
Proposition 1 can be cast in the form of an algorithm:
\begin{enumerate}
	\item  Compute the polynomial $p(t,T)=\det(c'(t),\dot d(T),d(T)-c(t))$.

	\item  Solve numerically (or analytically) $p(t_{i},T_{i})=0$ for $T_{i}$ for a sequence of values $0= 
	t_{0}<t_{1}<\cdots <t_{N-1}<t_{N}=1$.

	\item  Choose a real monotonically growing solution 
	$0=T_{0}<T_{1}<\cdots<T_{N-1}<T_{N}=1$.
	
	\item If there is no such solution, apply Rabl's method 
	\cite{rabl}, \cite{arribas} to the regions where the $T_{i}$ sequence is not 
	 growing.
	 
	 \item The developable surface patch is defined by the  
	 segments $\overline{c(t_{i})d(T_{i})}$.
\end{enumerate}

In the case of parabolas in parallel planes, the
reparameterization $T(t)$ is a M\"obius transformation. It is 
monotonically increasing unless its denominator vanishes. 
\begin{example}Construction of a developable surface patch bounded by 
curves $c$ and $d$ with respective control polygons \[\{(0,0,0), 
(0,1,0),(2,1,0)\}, \qquad\{(0,0,1), (0,3/2,1), (1,3/2,1)\}.\]
\end{example}

These are parabolas on parallel planes $z=0$, $z=1$,
\[c(t)=(2t^2, 2t-t^2, 0), \qquad d(T)=\left(T^2, 3T-\frac{3}{2}T^2, 
1\right).\]

Condition (\ref{algdev}) is just
\[8t T  + 4 T - 12 t=0\Rightarrow T(t)=  \frac{3 t}{2t+1},\quad 
t\in[0,1].\]

\begin{figure}
\begin{center}
    \includegraphics[height=0.2\textheight]{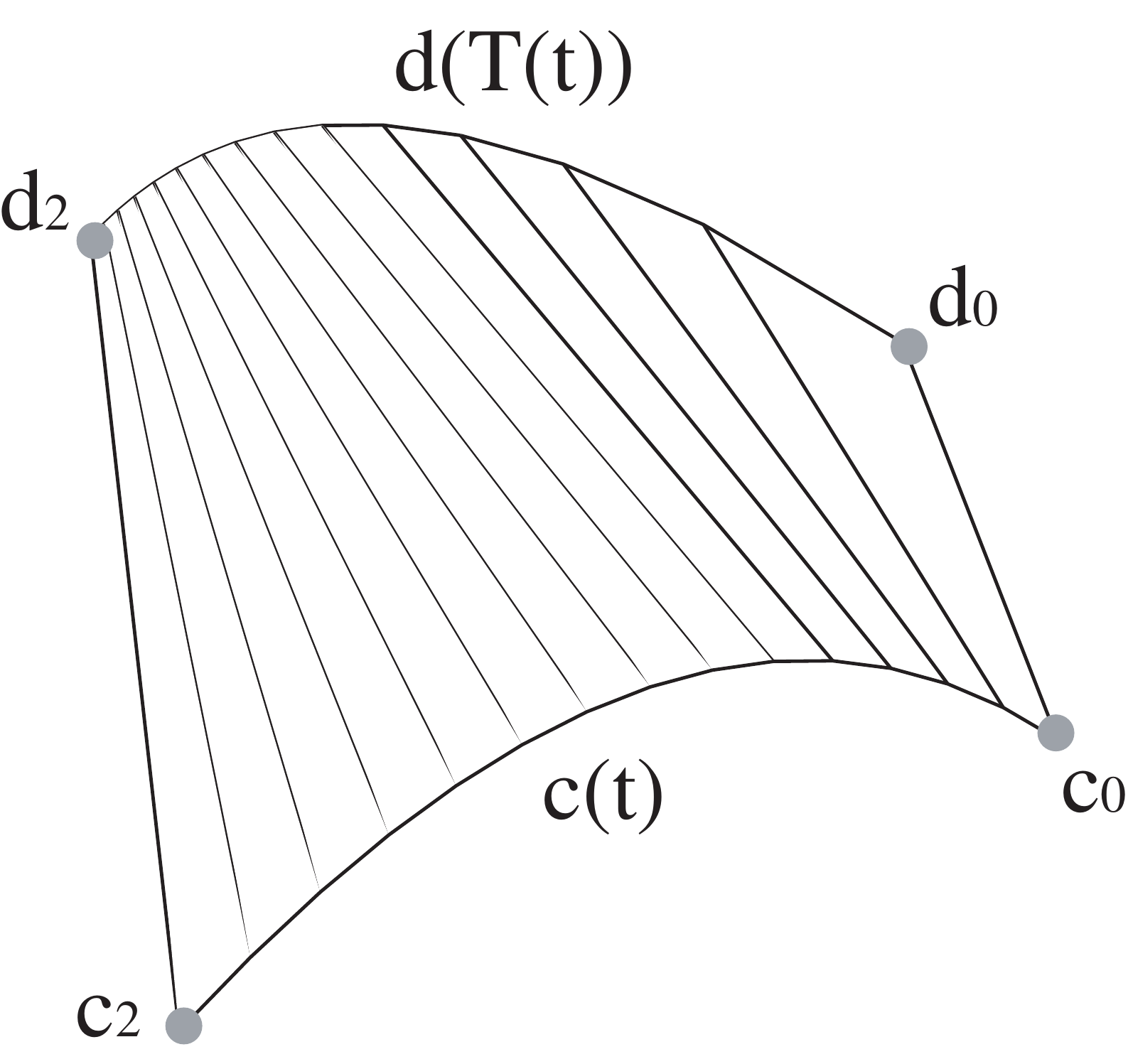}
\end{center}
\caption{Developable surface patch bounded by parabolas in parallel 
planes\label{para-planas}}
\end{figure}

Since $T(t)$ becomes singular just at $t=-1/2$, out of our patch, 
the parameterization
\[b(t,v)=(1-v)(2t^2, 2t-t^2, 0)+ v\left(T(t)^2, 3T(t)-\frac{3}{2}T(t)^2, 
1\right),\quad t,v\in[0,1],\]
corresponds to a developable surface patch bounded by $c$ and $d$ 
(See Fig.~\ref{para-planas}), which does not meet the edge of 
regression of the surface.

For cubic curves in parallel planes,  equation (\ref{algdev}) is of
degree two:

\begin{example}Construction of a developable surface patch bounded by 
cubic curves $c$ and $d$ with respective control polygons \[\{(0,0,0), (1,0,0),(2,1,0), 
(2,3,0)\}, \ \{(0,0,1), (3/2,0,1), (2,3/2,1), (2,5/2,1)\}.\]
\end{example}

These are cubic curves on parallel planes $z=0$, $z=1$,
\[c(t)=(-t^3+3t, 3t^2, 0), \qquad
d(T)=\left(\frac{1}{2}T^3-3T^2+\frac{9}{2}T, -2T^3+\frac{9}{2}T^2,
1\right).\]

\begin{figure}
\begin{center}
    \includegraphics[height=0.2\textheight]{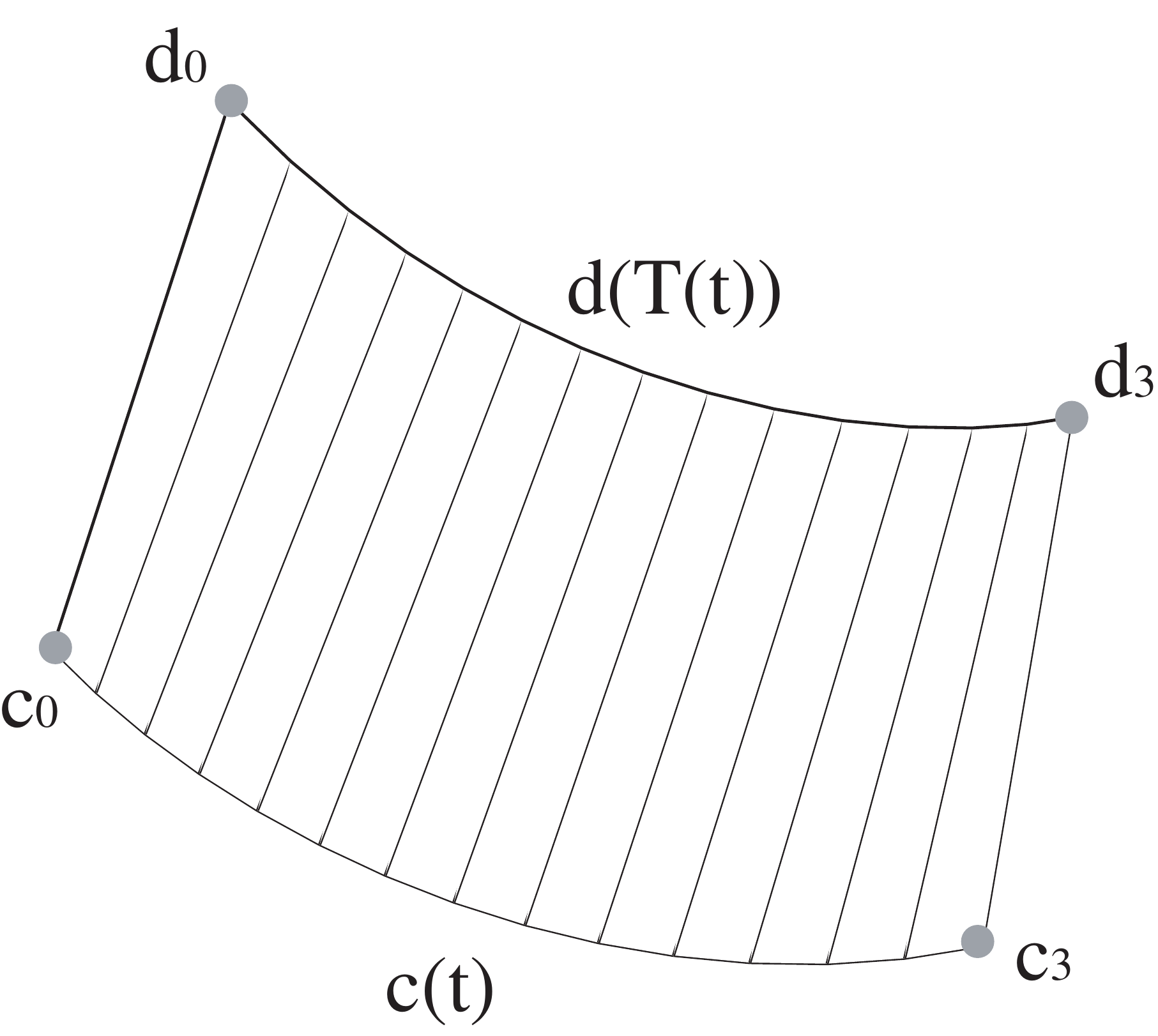}
\end{center}
\caption{Developable surface patch bounded by cubics on parallel 
planes\label{cubi-planas}}
\end{figure}

The developability condition (\ref{algdev}) 
\[(18+9t-18t^2)T^2+(27t^2-36t-27)T+27t=0\] has just one monotonically 
increasing solution mapping the interval $[0,1]$ onto itself,
\[T(t)=\frac{1}{2}\frac {3\,{t}^{2}-4\,t-3+\sqrt {9\,{t}^{4}-14\,{t}^{2}+9}}{2\,{t
}^{2}-t-2},\]
which produces a developable surface patch (See Fig.~\ref{cubi-planas}).

It is interesting to show what happens when this construction is
applied to a B\'ezier developable patch, that is, a polynomially 
parameterized
surface patch which is developable and which can be obtained, for 
instance, resorting to Aumann's construction \cite{aumann}:
\begin{example} Construction of a developable surface patch bounded by 
 cubic curves $c$ and $d$ with respective control polygons 
\[\{(0, 0, 0), (3, 3, 0), (4, 3, 0),(5,0,0)\}, \ \{(0, 0, 2), (2, 2, 
3), (13/6, 3/2, 9/2), (23/12, -5/4, 27/4)\},\]
\end{example}
and respective parameterizations
\[c(t)=(2t^3-6t^2+9t, -9t^2+9t, 0), \]\[
d(T)=\left(\frac{17}{12}T^3-\frac{11}{2}T^2+6T, \frac{1}{4}T^3-\frac{15}{2}T^2
+6T, \frac{1}{4}T^3+\frac{3}{2}T^2+3T+2\right).\]

The developability condition (\ref{algdev}) is a quartic equation
\begin{eqnarray*}-\frac{9}{4}(T+2)^2(T-t)
    ((6t^2+16t-5)T+6t^3-12t^2-17t-8)=0,
\end{eqnarray*}
which can be easily factored.

\begin{figure}
\begin{center}
    \includegraphics[height=0.15\textheight]{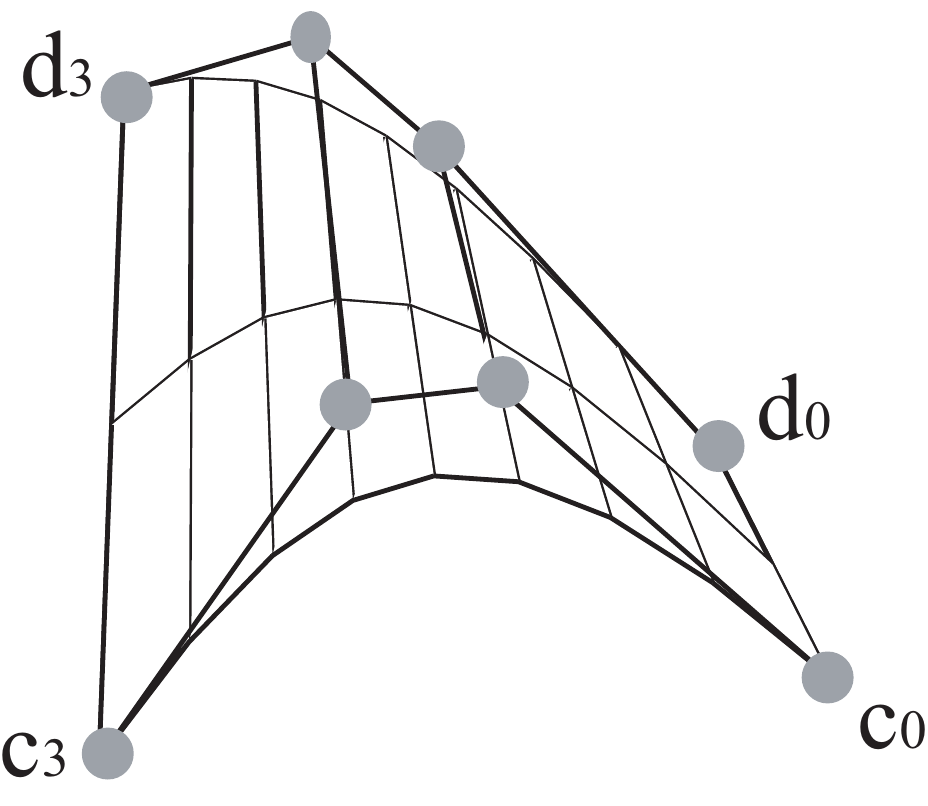}
\end{center}
\caption{B\'ezier developable surface patch bounded by cubics\label{aumann1}}
\end{figure}

The only monotonically increasing solution is the obvious one, 
$T(t)=t$. The other one is singular in the interval $[0,1]$. The 
surface patch may be seen in Fig.~\ref{aumann1}.

Now we show an example of a developable surface patch with a 
regression area: 

\begin{example}Construction of a developable surface patch bounded by 
quartic curves $c$ and $d$ with respective control polygons 
\[\{(0,0,0), (1,1,0), (2,1,0), (3,1,0), (4,0,0)\}, \quad
\{(0,0,1), (1,1,1), (2,-1/2,1), (3,1,1), (4,0,1)\}.\]
\end{example}
and respective parameterizations
\[c(t)=(4t, -2t^4+4t^3-6t^2+4t, 0), \quad
d(T)=\left(4T, -11T^4+22T^3-15T^2+4T, 1\right).\]

\begin{figure}
\begin{center}
    \includegraphics[height=0.3\textheight]{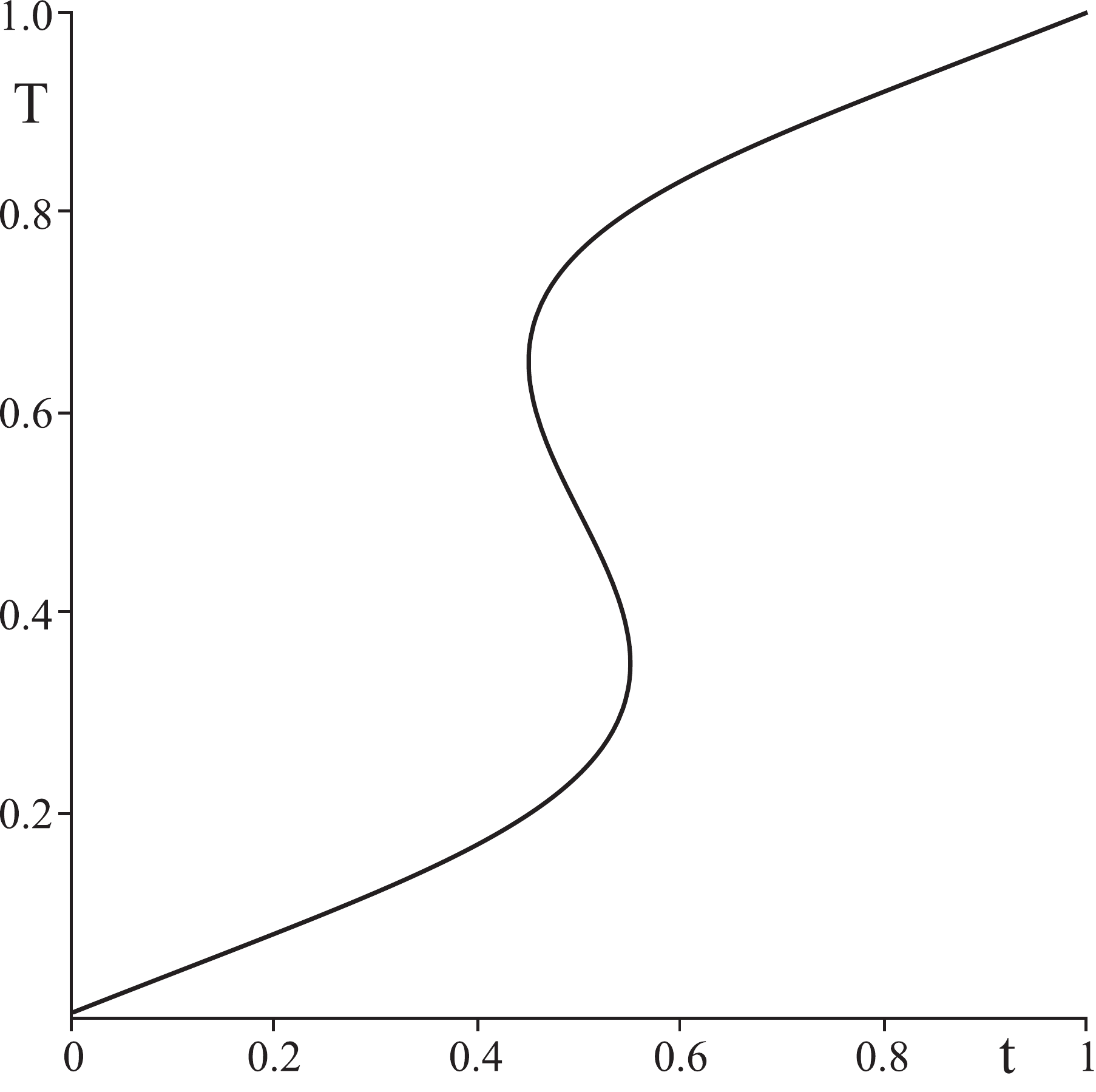}\hspace{1cm}
    \includegraphics[height=0.3\textheight]{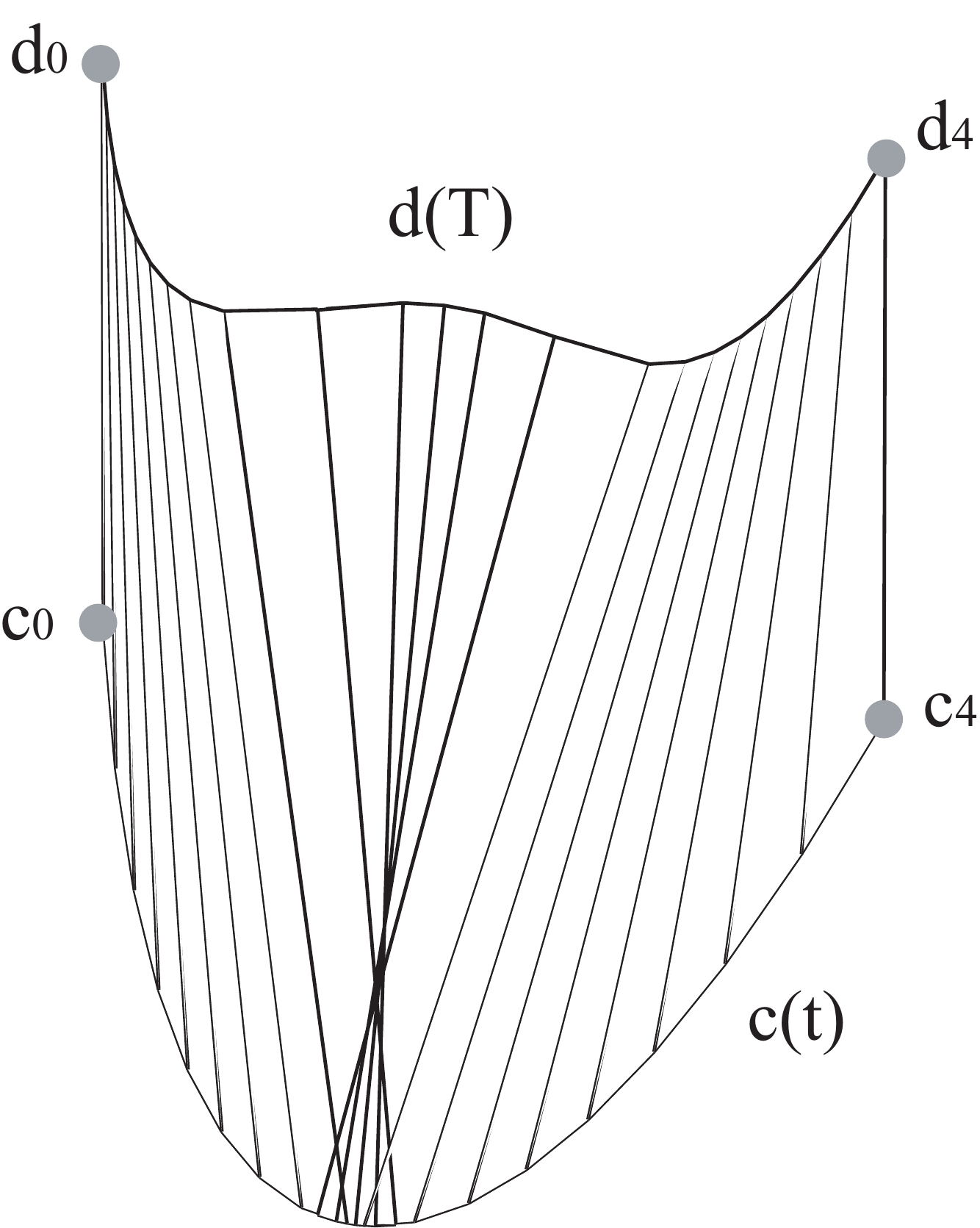}
\end{center}
\caption{Developable surface patch with a regression area\label{regres}}
\end{figure}

The developability condition (\ref{algdev}) is a quartic equation
\begin{eqnarray*}-176T^3-32t^3-264T^2+48t^2+120T-48t=0,
\end{eqnarray*}
and provides a reparameterization function $T(t)$  which is not 
monotonic and hence the surface patch has a regression area between 
$t=0.45$ and $t=0.55$, as it is seen in Fig.~\ref{regres}.

\begin{figure}
\begin{center}
    \includegraphics[height=0.25\textheight]{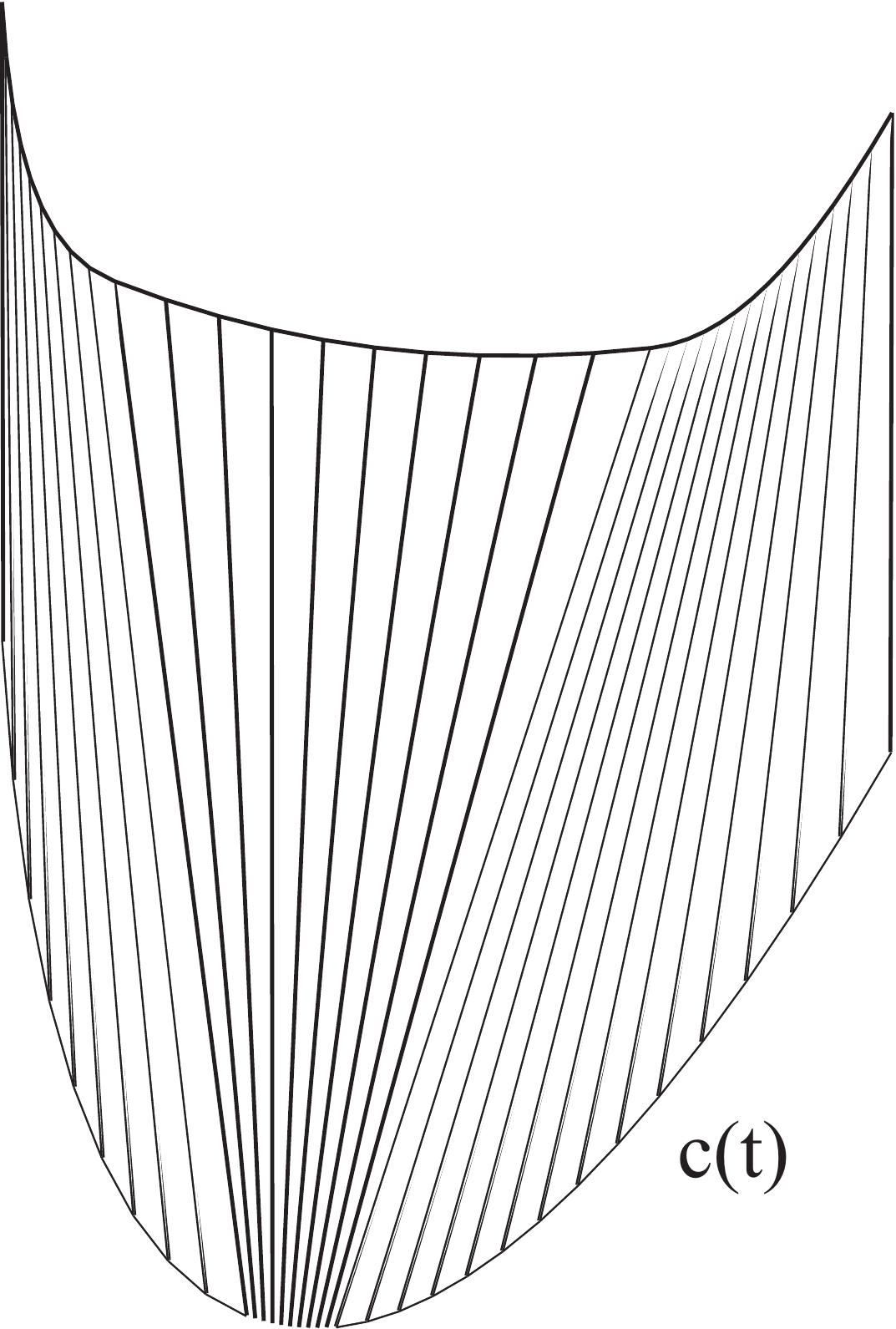}
\end{center}
\caption{Quasi-developable surface patch after amending the 
regression area\label{multic}}
\end{figure}

This is due to the bump in the graphic of the curve $d$ and can be 
amended (See Fig.~\ref{multic}) modifying this curve using the multiconic development 
\cite{rabl,arribas}.

A simple but realistic example of design of a vase:

\begin{example}Construction of a developable surface patch bounded by 
curves $c$ and $d$ with respective control polygons 
\[\{(1.05, 47.594, 79.90), (-3.697, 55.816, 59.541), (-0.225, 45.386, 
39.950)\}, \] \[\{(12.95, 47.594, 79.90), (14.711, 56.613, 59.401),(14.225, 45.386, 39.95)\}.\]\end{example}

Both curves $c(t)$, $d(t)$ are parabolas. Besides, we consider 
the straight segment with control polygon
\[(-4.9, 57.90, 79.90), (-7.45, 57.90, 39.95),\]
in order to construct an additional planar face bounded by $c(t)$ and 
the segment.

The developability condition (\ref{algdev}) is cubic,
\[(10408.24-16679.66t)T^2+(16700.06t^2+190.32t+19210.01)T-10254.48t^2-17870.03t-717.57=0,\]
and can be solved explicitly for $T(t)$,
\[T(t)=\frac { 1.67 t^{2}+ 0.019 t+ 1.92-\sqrt {2.79 t^{4}- 6.78 {t}^{3}-
 1.24 t^{2}+ 7.03 t+ 3.99}}{
3.32 t- 2.08}
.\]

\begin{figure}
\begin{center}
    \includegraphics[height=0.3\textheight]{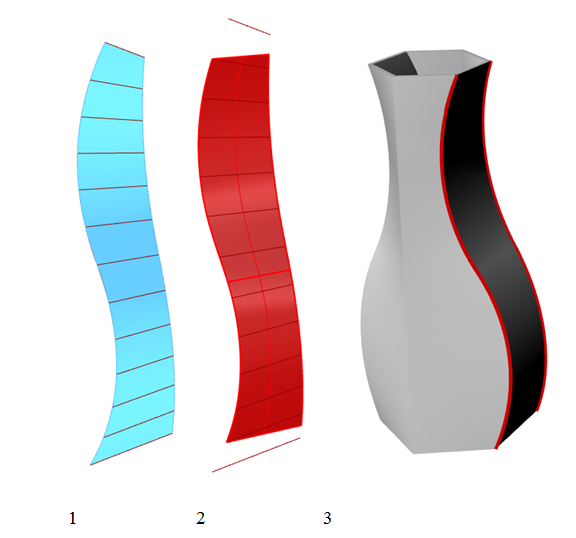}
    \includegraphics[height=0.3\textheight]{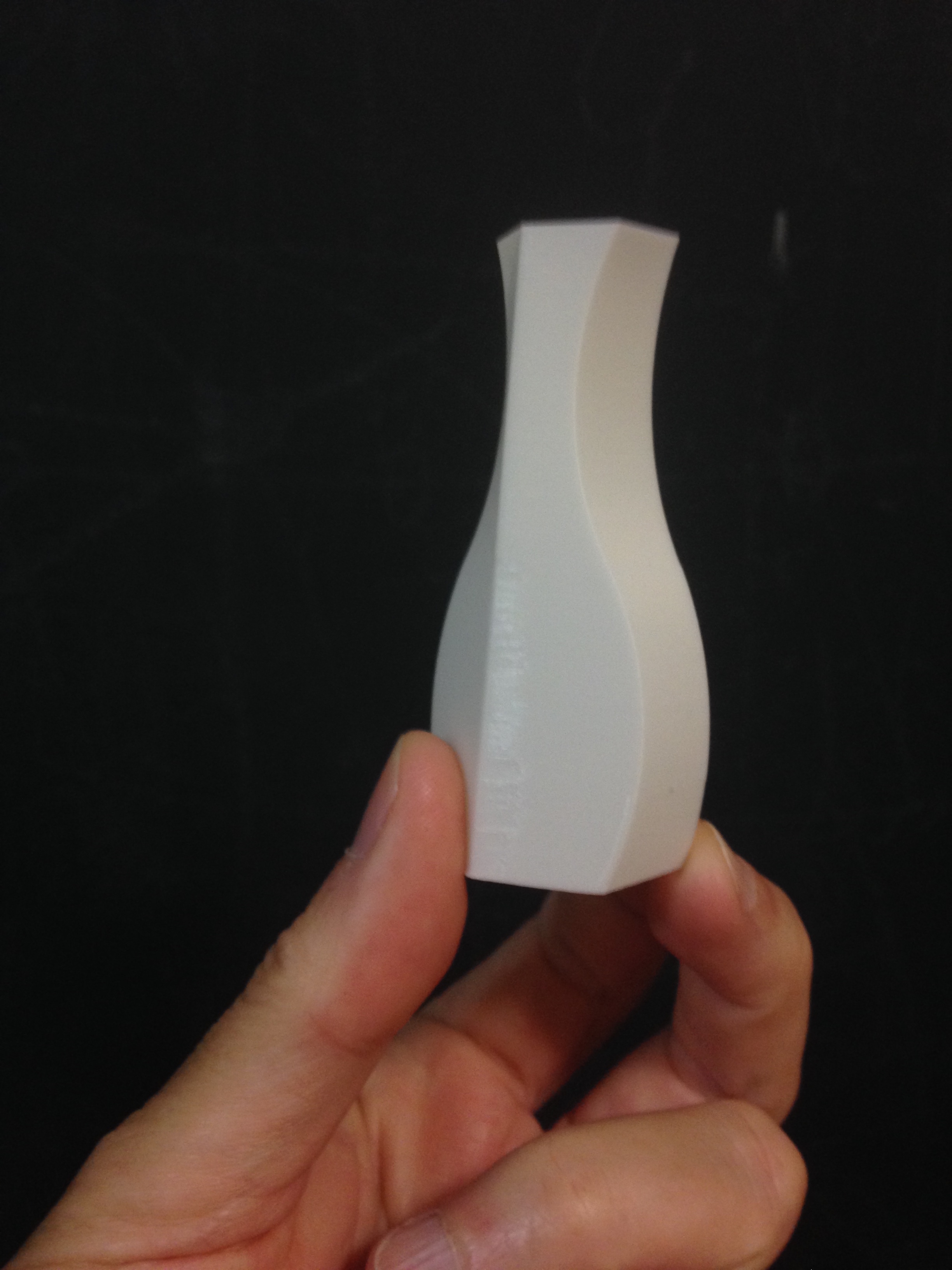}
\end{center}
\caption{Vase formed by devalopable patches\label{vase}}
\end{figure}

The result may be seen in Fig~\ref{vase}, where 1 shows the original
patch, duplicated in a symmetric fashion, 2 shows the trimmed patches
and 3 shows the final object after repetition and inclusion of 
planar faces. 

We finish this section with an example with cubic curves:
\begin{example}Construction of a developable surface patch bounded by 
rational cubic curves $c$ and $d$ with respective control polygons 
\[\{(0,0,0), (1,0,0),(2,1,0), (2,3,0)\}, \ \{(0,0,1), (3/2,0,3/2), 
(3/2,3/2,3/2),(3/2,5/2,3/2)\},\]and lists of weights,
\[\{1,1/2,1/3,1\}, \qquad \{1,1/3,1/4,1\},\]
\end{example}
which are parameterized as
\[c(t)=\left(\frac{t(3t^2-2t+3)}{t^3+2t^2-3t+2}, 
\frac{4t^3+2t^2}{t^3+2t^2-3t+2}, 0\right), \]\[
d(T)=\left(\frac{15T^3-15T^2+12T}{2T^3+14T^2-16T+8}, 
\frac{T^2(9+11T)}{2T^3+14T^2-16T+8}, 
\frac{7T^3+9T^2-12T+8}{2T^3+14T^2-16T+8}
\right).\]

\begin{figure}
\begin{center}
    \includegraphics[height=0.2\textheight]{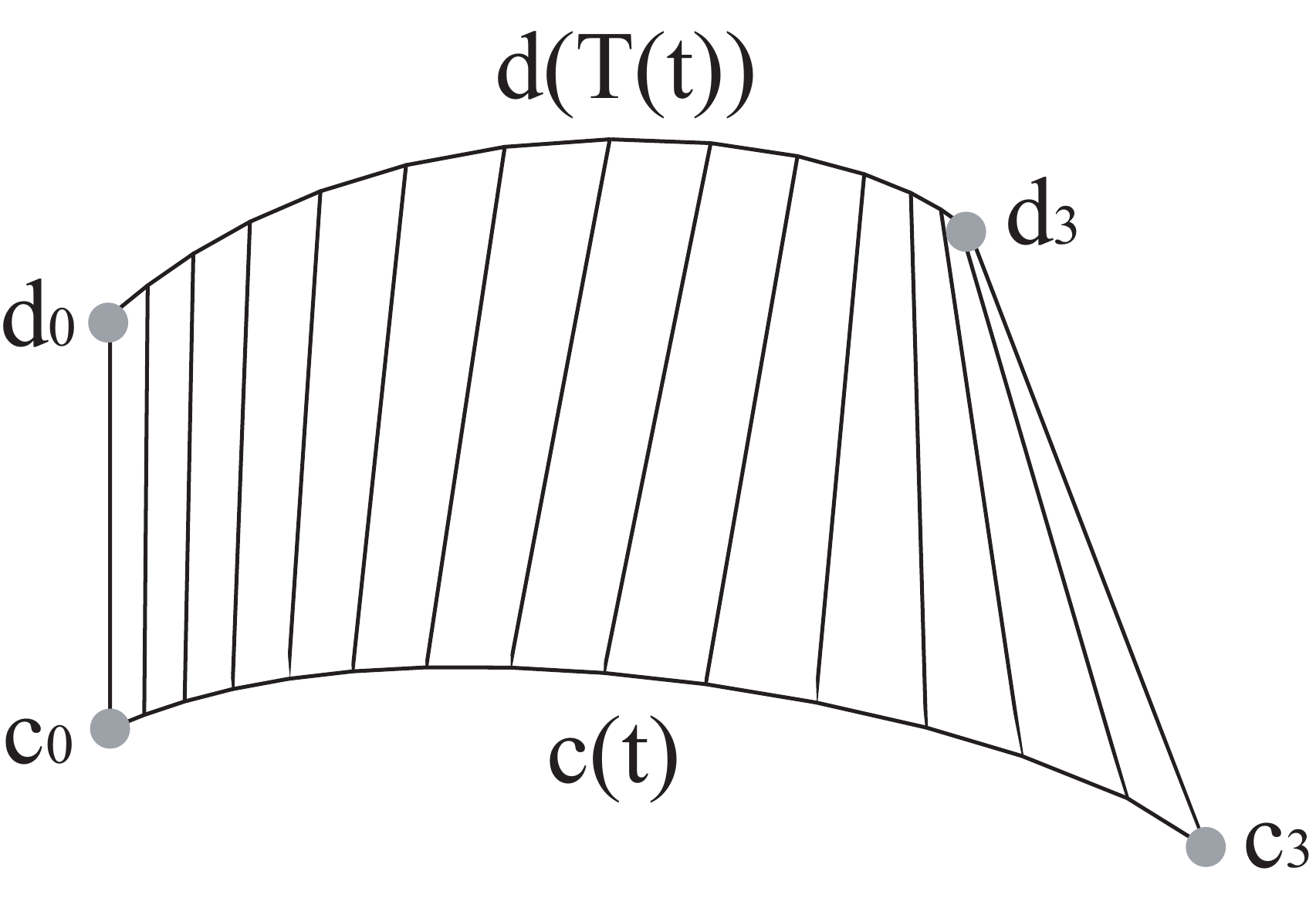}
\end{center}
\caption{Developable surface patch bounded by two rational cubics
\label{cubic}}
\end{figure}

In this case the developability condition (\ref{algdev}) 
\begin{eqnarray*}0&=&(4t^4-372t^3+519t^2+276t-27)T^4+
	(176t^4+264t^3-726t^2-792t+198)T^3
\\&+&(-84t^4-396t^3+729t^2+588t-117)T^2+(-184t^4+912t^3-924t^2-96t-108)T
\\&+&16t^4-192t^3+240t^2+96t
\end{eqnarray*}
is a quartic equation, which can be solved numerically as a function 
$T(t)$ interpolating 
a list of pairs of values for $t$ and $T$,
\begin{center}
\begin{tabular}{c||c|c|c|c|c|c|c|c|c|c|c|c}
T & 0.0 & 0.1 & 0.2 & 0.3 & 0.4 & 0.5 & 0.6 & 0.7 & 0.8 & 0.9 & 1.0 \\\hline
t & 0.0 & 0.091 &  0.1799 &  0.26.39 & 0.3440
&  0.4228 &  0.5036 & 0.5900 & 0.6863 & 0.7992 &1.0 \\
\end{tabular}
\end{center}
in order to achieve a developable surface patch (See Fig.~\ref{cubic}).

\section{Developable patches bounded by NURBS curves\label{devspline}}

The results for developable surface patches bounded by rational B\'ezier curves
can be trivially extended to rational spline curves of degree $n$, $c(t)$, $d(T)$,
$t,T\in[0,1]$, with respective lists of knots
$\{0<t_{1}\le\ldots \le t_{c}<1\}$, $\{0<T_{1}\le\ldots\le T_{d}<1\}$.  One has just to 
apply the algorithm in Section~\ref{examp} to the pieces of both 
curves:
\begin{enumerate}
	\item  Apply the algorithm to the first piece of $c(t)$ and 
	$d(T)$ till reaching $t_{1}$ or $T_{1}$:
	
\begin{enumerate}
	\item  If $t_{1}$ is reached, apply the algorithm to the second 
	piece of $c(t)$ and the first piece of $d(T)$.

	\item  If $T_{1}$ is reached, apply the algorithm to the first
	piece of $c(t)$ and the second piece of $d(T)$.
\end{enumerate}

\item Proceed in a similar way on reaching $t_{2}$, $T_{2}$,\dots, 
$t_{c}$, $T_{d}$.	 
	 \item The developable surface patch is defined by the  
	 segments $\overline{c(t_{i})d(T_{i})}$.
\end{enumerate}

As it has already been said, the piecewise function $T(t)$ belongs to 
the class $C^{k-1}$ of differenciability if the parameterized curves belong 
to the class $C^{k}$.
%
%
\begin{example}Construction of a developable surface patch bounded by 
curves $c$ and $d$ with respective B-spline polygons 
\[\{(0,0,0),(0,1,0),(1,3/2,0), (2,1,0),(3,0,0)\}, \quad
\{(0,0,1),(0,3/2,1),(1/2,2,1),(1,2,1),(2,1,1)\},\] and common list of 
knots $\{0,0,0,1,2,2,2\}$. Both curves are cubics of 
two pieces.
\end{example}

For the interval $[0,1/2]$ we have parameterizations for both curves,
\[c_{1}(t)=\left(\frac{-t^3+3t^2}{2}, 
\frac{t^3}{2}-\frac{9t^2}{4}+3t, 0\right),\quad
d_{1}(T)=
\left(-\frac{T^3}{4}+\frac{3T^2}{4}, 
\frac{9T^3}{8}-\frac{15T^2}{4}+\frac{9T}{2}, 1\right)\]

On the other hand, the parameterizations on  the interval $[1/2,1]$ are,
\[c_{2}(t)=\left(\frac{t^3}{2}-\frac{3t^2}{2}+3t-1, 
-\frac{t^3}{2}+\frac{3t^2}{4}+1, 0\right),\quad
d_{2}(T)=
\left(\frac{3T^3}{4}-\frac{9T^2}{4}+3T-1, 
-\frac{7T^3}{8}+\frac{9T^2}{4}-\frac{3T}{2}+2, 1\right).\]

The developability condition (\ref{algdev})  for $c_{1}(t)$, 
$d_{1}(T)$,
\[\left(\frac{63}{16}t^2-\frac{27}{4}t-\frac{9}{4}\right)T^2+
\left(-9t^2+\frac{63}{4}t+\frac{9}{2}\right)T+
\frac{27}{4}t^2-\frac{27}{2}t=0,\]
has a monotonically growing solution
\[T_{1}(t)=\frac{2(4t+1-\sqrt {-5t^{2}+2\,t+1})}{7t+2},
\] which reaches the value $T=1$ for $t=2/3$.  

For $t\in[2/3,1]$, $T\in[1,2]$, we have to apply the developability
condition (\ref{algdev}) to $c_{1}(t)$, $d_{2}(T)$,
\[\left(-\frac{9}{16}t^2-\frac{9}{4}t+\frac{27}{4}\right)T^2+
\left(\frac{27}{4}t-\frac{27}{2}\right)T
+\frac{9}{4}t^2-9t+9=0,\]
which has a monotonically increasing solution,
\[T_{2}(t)=\frac{2(3+\sqrt{t^2+4t-3})}{t+6},\]
which for $t=1$ reaches the value $T=(6+2\sqrt{2})/7$.

Finally, for $t\in[1,2]$, $T\in[(6+2\sqrt{2})/7,2]$, we apply the developability
condition (\ref{algdev}) to $c_{2}(t)$, $d_{2}(T)$,
\[\left(\frac{9}{16}t^2-\frac{9}{2}t+\frac{63}{8}\right)T^2+
\left(\frac{27}{4}t-\frac{27}{2}\right)T
-\frac{9}{4}t^2+\frac{9}{2}=0,\]
for which the monotonically increasing solution is 
\[T_{3}(t)=\frac{2(-3t+6+\sqrt{t^4-8t^3+21t^2-20t+8})}{t^2-8t+14},\]
which reaches $T=2$ at $t=2$.

Hence, the list of knots for $c(t)$ has to refined to 
$\{0,0,0,2/3,1,2,2,2\}$ and the one for $d(T)$ to 
$\{0,0,0,1,(6+2\sqrt{2})/7,2,2,2\}$. 

\begin{figure}
\begin{center}
    \includegraphics[height=0.2\textheight]{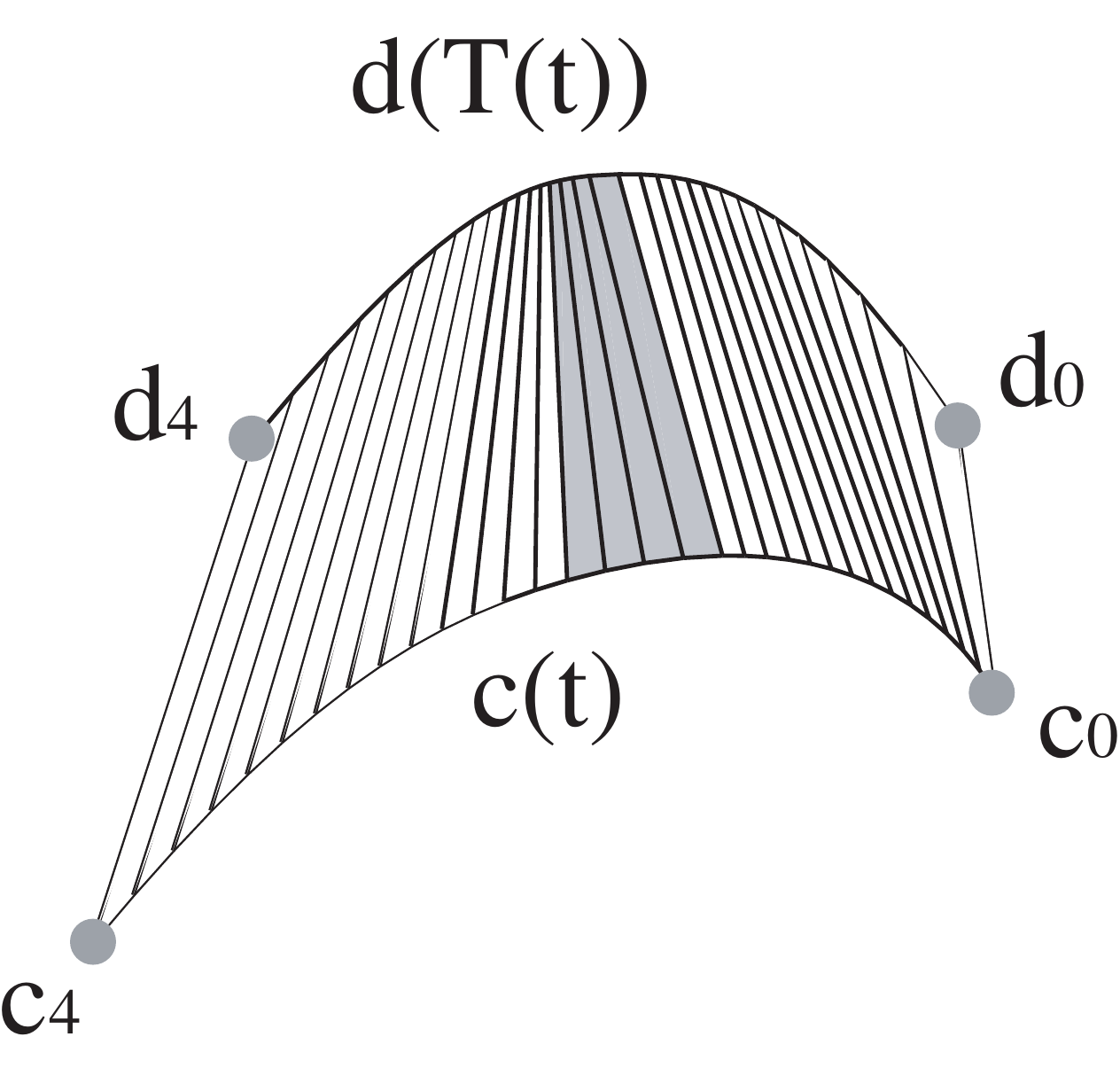}
\end{center}
\caption{Developable surface patch bounded by spline cubics 
\label{spline-cubic}}
\end{figure}

The parameterized surface patch (See Fig.~\ref{spline-cubic})
is developable for
\[T(t)=\left\{
\begin{array}{ll} T_{1}(t) & t\in[0,2/3]  \\
T_{2}(t) & t\in[2/3,1]  \\
T_{1}(t) & t\in[1,2].  \\\end{array}\right.\]

It can be easily checked that the piecewise function $T(t)$ belongs 
to the class $C^1$ of differenciability, as expected, since cubic splines 
with simple knots belong to the class $C^2$.

\begin{example}This example is borrowed from \cite{chalfant} and 
corresponds to the hull of a boat : We have three curves named 
sheer, chine and center line, with respective B-spline polygons 
\[ c_{0}=(0.00, 0.00, 9.00),\ 
c_{1}=(6.86, 7.10, 8.22), \ 
c_{2}=(21.6, 8.93, 6.25), \]\[
c_{3}=(36.9, 8.73, 5.86), \
c_{4}=(45.0, 7.65, 6.10),\]
\[ d_{0}=(1.40,0.00,5.30),\ 
 d_{1}=(10.5,7.53,1.93),\ 
 d_{2}=(25.7,7.85,1.28),\]\[ 
 d_{3}=(40.4,7.46,1.27),\ 
 d_{4}=(44.1, 7.20, 1.70),\]
\[ e_{0}=(1.40, 0.00, 5.30),\ 
e_{1}=(2.26,0.00, -0.21),\ 
e_{2}=(22.6,0.00, -0.10),\]\[ 
e_{3}=(36.3,0.00, -0.10),\ 
e_{4}=(44.1, 0.00, 0.50).\]
The list of knots is $\{0,0,0,1,2,2,2\}$ for the three curves and
hence they all are cubic splines of two pieces.
\end{example}

We construct two developable surface patches,
bounded by these curves, in order to have a developable hull.

The values of the parameters corresponding to the endpoints of each 
ruling can be seen in Table 1.
%
\begin{table}\hspace{-4cm}\scriptsize
\begin{tabular}{c|c||c|c|c|c|c|c|c|c|c|c|c|c|c|c|c|c|c|c|c|c|c|c|c|c|c|c|}
$t_{d}$ & 0.0  &  0.1 & 0.2 & 0.3 & 0.4 & 0.5 & 0.6 & 0.7 & 0.8 & 0.9
& 1.0  &  1.1 & 1.2 & 1.3 & 1.4 & 1.5 & 1.6 & 1.7 & 1.8 & 1.9 & 2.0\\
\hline
$T_{c}$ &0.06    &  0.15  &  0.25  &  0.35  &   0.46  &    0.58  & 
0.71  &   0.85  &   1.09  &   1.29  &  1.38  &   1.40  &   1.43  &  1.46  &
  1.49  &  1.52  &    1.56  &   1.60 & 1.65  &   1.71  &   1.79   \\
\hline
$T_{e}$ & 0.00   &  0.09   &  0.17 &  0.26  &  0.35   &  0.44   &  0.53  &  0.61 
 &  0.69   &  0.76  &  0.80  &  0.82   &  0.85  &  0.90   &  1.19  &  1.41 
 &  1.56  &  1.71 &  1.88   &  2.08   &  2.39 
\end{tabular}
\caption{Values of $t, T$ on the curves $c(T)$, $d(t)$, $e(T)$ that determine the 
rulings of the developable hull}
\end{table}


The curves have been chosen 
without fulfilling the coplanarity requirement at $t=0$ and at $t=2$, 
since the segments $\overline{c_{0}d_{0}}$, $\overline{c_{4}d_{4}}$, 
$\overline{e_{4}d_{4}}$ are not rulings of a developable surface. For 
this reason we are to shorten or enlarge the boundary curves.

\begin{figure}
\begin{center}
    \includegraphics[height=0.2\textheight]{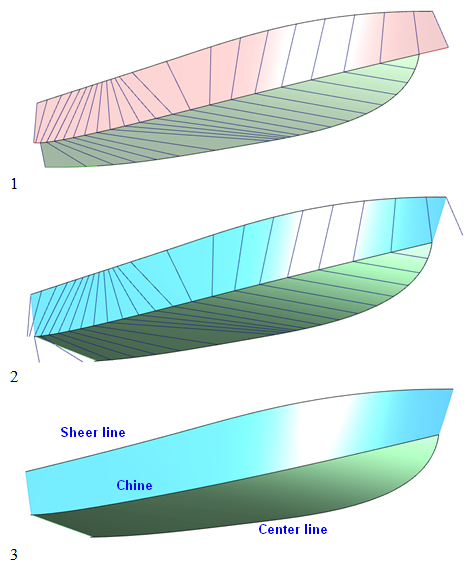}
    \includegraphics[height=0.2\textheight]{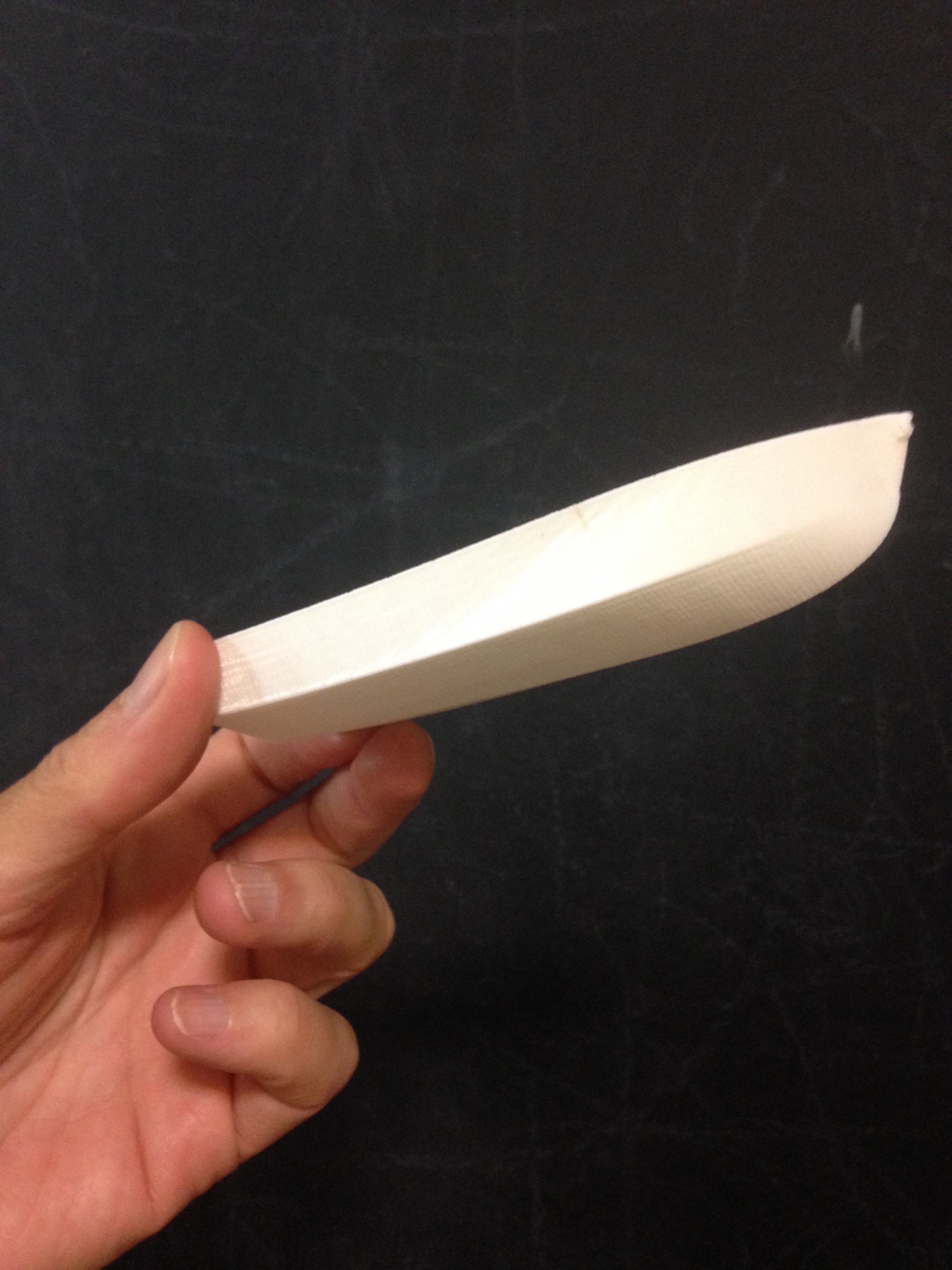}
\end{center}
\caption{Developable surface patches bounded by the sheer, chine and 
base of the hull of a boat \label{hull}}
\end{figure}

The resulting developable surfaces may be seen in Fig.~\ref{hull},
showing (1) the lofting surface of the rulings between the extended
boundary curves, (2) the trimmed surface at the ends and (3) the final
representation of the hull.

\section{Conclusions}

In this paper we propose a simple method for constructing developable
surface patches bounded by two rational B\'ezier or rational spline
curves.  The method is founded on finding a reparameterization function
for one of the boundary curves.  The most relevant feature of this
construction is that the equation for this function is algebraic and
of low degree and it is amenable to solving by simple numerical
methods, or even analytical methods, since the highest possible degree
for cubic curves is four.  The requirement of monotonicity for the
reparameterization function is seen to be simple in terms of the
accelerations or curvature vectors of the bounding curves.

\section*{Acknowledgments}
 
This work is partially supported by the Spanish Ministerio de
Econom\'\i a y Competitividad through research grant TRA2015-67788-P.

\bibliographystyle{elsarticle-num}
\bibliography{cagd}

\end{document}